\documentclass[11pt]{article}

\usepackage[paper=a4paper,dvips,top=2.0cm,left=2.2cm,right=2.2cm,bottom=3.2cm,footskip=1.6cm]{geometry}
\usepackage{amsfonts,amsmath,amsthm,amssymb}
\numberwithin{equation}{section}  
\usepackage[svgnames]{xcolor}
\usepackage{fix-cm}
\usepackage{sectsty}
\usepackage{fancyhdr}
\usepackage{lastpage}
\usepackage{graphicx}
\usepackage{url}
\usepackage{enumerate}
\usepackage{paralist}
\usepackage{multicol}
\usepackage{color}  
\usepackage{float}  
\usepackage{epsfig}
\usepackage{hyperref}   
\hypersetup{colorlinks=true,linkcolor=beamer@PRD, citecolor=beamer@PRD}
\usepackage{authblk}  
\usepackage{cite}  

\definecolor{beamer@blue}{RGB}{0,0,255}
\definecolor{beamer@mediumblue}{RGB}{0,0,190}
\definecolor{beamer@midnightblue}{RGB}{25,25,112}
\definecolor{beamer@navy}{RGB}{0,0,128}
\definecolor{beamer@darkblue}{RGB}{0,0,139}
\definecolor{beamer@purple}{RGB}{128,0,128}
\definecolor{beamer@levander}{RGB}{100.,149.,237.}
\definecolor{beamer@PRD}{RGB}{46,48,146}
\definecolor{beamer@green}{RGB}{0,128,0}
\definecolor{beamer@darkgreen}{RGB}{0,100,0}
\definecolor{beamer@olive}{RGB}{128,128,0}
\definecolor{beamer@darkolivegreen}{RGB}{85,107,47}
\definecolor{beamer@gray}{RGB}{190,190,190}
\definecolor{beamer@ivry}{RGB}{220,220,220}
\definecolor{beamer@new}{RGB}{40,120,50}
\definecolor{shadecolor}{RGB}{220,220,220}
\definecolor{beamer@darkslategray}{RGB}{47,79,79}
\definecolor{beamer@chocolate}{RGB}{210,105,30}
\definecolor{beamer@brown}{RGB}{165,42,42}
\definecolor{beamer@orangered}{RGB}{255,69,0}
\definecolor{beamer@maroon}{RGB}{128,0,0}
\definecolor{beamer@white}{RGB}{234,242,243}
\definecolor{beamer@silver}{RGB}{0.5,0.45,0.37}

\begin{document}


\title{\textbf{Entangled squeezed states in noncommutative spaces with minimal length uncertainty relations}}


\author{\textbf{Sanjib Dey$^{1,3}$ and V{\'e}ronique Hussin$^{1,2}$}}
\date{\footnotesize{$^{1}$Centre de Recherches Math{\'e}matiques (CRM), Universit{\'e} de Montr{\'e}al, Montr{\'e}al - H3C 3J7, Qu{\'e}bec, Canada \\ $^{2}$Department de Math{\'e}matiques et de Statistique, Universit{\'e} de Montr{\'e}al, Montr{\'e}al - H3C 3J7, Qu{\'e}bec, Canada \\ $^{3}$Department of Mathematics and Statistics, Concordia University, Montr{\'e}al - H3G 1M8, Qu{\'e}bec, Canada \\ \small{E-mail: dey@crm.umontreal.ca, veronique.hussin@umontreal.ca}}}
\maketitle
    	
\thispagestyle{fancy}
\begin{abstract}
We provide an explicit construction of entangled states in a noncommutative space with nonclassical states, particularly with the squeezed states. Noncommutative systems are found to be more entangled than the usual quantum mechanical systems. The noncommutative parameter provides an additional degree of freedom in the construction by which one can raise the entanglement of the noncommutative systems to fairly higher values beyond the usual systems. Despite of having classical-like behaviour, coherent states in noncommutative space produce little amount of entanglement and therefore they possess slight nonclassicality as well, which are not true for the coherent states of ordinary harmonic oscillator. 
\end{abstract}	 
\begin{section}{Introduction} \label{sec1}
\addtolength{\voffset}{1.2cm} 
Entanglement or nonseparability of states is probably the most fascinating property of quantum systems, which was elucidated by Einstein and his collaborators in their famous EPR-paradox \cite{Einstein_Podolsky_Rosen}. A possible resolution of the paradox was provided by Einstein in terms of the hidden variable theory, and therefore, he claimed that the conventional quantum mechanics was incomplete. Later Bell's inequality \cite{Bell} was shown to be violated for many systems, and thus, the hidden variable theory failed to explain the paradox, which tends to confirm that the original formulation of quantum mechanics is indeed correct. However, in spite of its paradoxical nature, entanglement has been detected experimentally and is recognised as a source of many important observations in quantum information theory including quantum teleportation \cite{Bouwmeester_Pan_Mattle}, quantum dense coding \cite{Mattle_Weinfurter_Kwiat_Zeilinger}, quantum cryptography \cite{Bennett_Bessette} and many more. Various devices have been proposed and realized experimentally to generate quantum entanglement, such as beam splitters \cite{Fearn_Loudon,Tan_Walls_Collett,Scheel_Welsch}, cavity QED \cite{Imamog}, NMR \cite{Gershenfeld_Chuang}, semiconductor microcavity \cite{Eleuch} etc.

Nonclassical effects are useful in quantum information theory. Squeezed states are highly nonclassical \cite{Walls,Loudon_Knight}. However, sometimes they are difficult to generate, as there is no generalised setting available in the literature to construct them. Additional complications arise when one considers the underlying space-time structure to be noncommutative, where the space-time coordinates do not commute any more. The most commonly studied version of these space-time structures consists of replacing the standard set of commutation relations for the
canonical coordinates $x^\mu$ by noncommutative versions, such as $[x^\mu,x^\nu]=i\hbar \theta^{\mu\nu}$, where $\theta^{\mu\nu}$ is taken to be a constant antisymmetric tensor. More interesting structures, leading for instance to minimal length and generalized versions of Heisenberg's uncertainty relations, are obtained when $\theta^{\mu\nu}$ is taken to be a function of the momenta and coordinates, e.g. \cite{Kempf,Bagchi_Fring,Fring_Gouba_Scholtz,Fring_Gouba_Bagchi,Dey_Fring_Gouba,Dey_Fring_Khantoul}. 

In the present work, we utilize the latter version of the noncommutative space-time and construct the squeezed states of a perturbed harmonic oscillator in this space. Our principal motivation is to construct the entangled states in noncommutative space, which have not been studied significantly to our knowledge. We employ the beam splitter for this purpose and transmit the noncommutative squeezed states through one of its input channels, while a vacuum state through the other. We analyse the density matrix of the output states to measure the linear entropy. The outcomes are astonishing, when we compare them with the similar computations coming out of the usual spaces. We always obtain higher entanglement in noncommutative spaces than the ordinary systems. The most interesting observation is that when we switch on the noncommutative parameter and increase the noncommutativity further, the entanglement rises correspondingly. This leads us to the possibility that we may use the noncommutative systems for the purpose of quantum information processing to obtain better results than the ordinary cases. One obvious question that arises afterwards, are the noncommutative systems accessible physically? The answer is yes \cite{Kempf,Bagchi_Fring,Fring_Gouba_Scholtz,Fring_Gouba_Bagchi,Dey_Fring_Gouba,Dey_Fring_Khantoul}, which has been clarified in section \ref{sec3}.  

Our manuscript is organised as follows: In section \ref{sec2}, we provide a general construction procedure, which can be used to build up the squeezed coherent states for any generalised models as well as for the deformed quantum systems. In section \ref{sec3}, we assemble various generalities on squeezed coherent states from section \ref{sec2} and construct the nonlinear coherent states and the squeezed states for a specific harmonic oscillator in a noncommutative space \cite{Kempf,Bagchi_Fring,Fring_Gouba_Scholtz,Fring_Gouba_Bagchi,Dey_Fring_Gouba,Dey_Fring_Khantoul}. In section \ref{sec4}, we measure the quantum entanglement of noncommutative systems by computing the linear entropy of the corresponding models and provide their comparative analysis with the ordinary quantum systems. Our conclusions are stated in section \ref{sec5}. 
\end{section}
\begin{section}{Generalised squeezed coherent states}\label{sec2}
We commence by discussing the basic notions of squeezed states. Squeezed states are obtained by applying the Glauber's unitary displacement operator $D(\alpha$) on the squeezed vacuum \cite{Nieto_Truax_PRL}
\lhead{Entangled squeezed states in noncommutative spaces}
\chead{}
\rhead{}
\begin{equation}\label{SqOperator}
\vert \alpha,\zeta\rangle=D(\alpha)S(\zeta)\vert 0\rangle, \qquad S(\zeta)=e^{\frac{1}{2}(\zeta a^\dagger a^\dagger-\zeta^\ast aa)}, \qquad D(\alpha)=e^{\alpha a^\dagger-\alpha^\ast a}, \qquad \alpha, \zeta \in \mathbb{C},
\end{equation}
with $\alpha$, $\zeta$ being displacement and squeezing parameters, respectively, and $S(\zeta)$ being the unitary squeezing operator. The ordering of $D(\alpha)S(\zeta)$ and $S(\zeta)D(\alpha)$ in \textcolor{beamer@PRD}{(}\ref{SqOperator}\textcolor{beamer@PRD}{)} are equivalent, amounting to a change of parameter \cite{Nieto_Truax_PRL}. An alternative ladder operator definition of the squeezed states can be obtained by performing the Holstein-Primakoff / Bogoliubov transformation on the squeezing operator \cite{Nieto_Truax,Nieto_Truax_PRL}. The squeezed states $\vert \alpha,\zeta\rangle$ can be constructed from the solution of the equivalent ladder operator definition as follows \cite{Fu_Sasaki,Alvarez_Hussin}
\begin{equation}\label{eigen}
(a+\zeta a^\dagger) \vert \alpha,\zeta\rangle=\alpha \vert \alpha,\zeta\rangle, \qquad \alpha,\zeta \in \mathbb{C}~.
\end{equation}
The coherent states are the special solutions when $\zeta=0$. A direct generalisation of the above definition is carried out by replacing the bosonic creation and annihilation operators $a, a^\dagger$ by a couple of generalised ladder operators $A, A^\dagger$ \cite{Alvarez_Hussin,Angelova_Hertz_Hussin}, such that
\begin{equation} \label{GenLad}
A^\dagger \vert n\rangle = \sqrt{k(n+1)} \vert n+1\rangle, \qquad A \vert n\rangle = \sqrt{k(n)} \vert n-1\rangle,
\end{equation}  
where $k(n)$ is fairly a general function leading to different generalised models. A second approach of generalising the ladder operators can be realized in terms of the $f$-oscillators \cite{Manko_Marmo_Solimeno_Zaccaria}
\begin{equation} \label{qOsc}
A_f^\dagger = f(n) a^\dagger, \qquad A_f = a f(n),
\end{equation}
where $f(n)$ is an operator-valued function of the Hermitian number operator $n=a^\dagger a$. The two different approaches \textcolor{beamer@PRD}{(}\ref{GenLad}\textcolor{beamer@PRD}{)} and \textcolor{beamer@PRD}{(}\ref{qOsc}\textcolor{beamer@PRD}{)} become equivalent with the choice $k(n)=n f^2(n)$. For clarifications on physical reality of the ladder operators $A, A^\dagger$ and $A_f, A_f^\dagger$, we refer the readers to \cite{Dey_Fring_Gouba_Castro,Dey}, where an explicit Hermitian representation of the operators was found in terms of the usual canonical observables. A natural checkpoint is the limit $f(n)=1$, where the operators in \textcolor{beamer@PRD}{(}\ref{GenLad}\textcolor{beamer@PRD}{)} and \textcolor{beamer@PRD}{(}\ref{qOsc}\textcolor{beamer@PRD}{)} reduce to the usual creation and annihilation operators as expected. In order to solve the eigenvalue equation \textcolor{beamer@PRD}{(}\ref{eigen}\textcolor{beamer@PRD}{)} for the generalised case, let us expand the squeezed states $\vert \alpha,\zeta\rangle$ in terms of the Fock states
\begin{equation}\label{Expansion}
\vert \alpha,\zeta\rangle=\frac{1}{\mathcal{N}(\alpha,\zeta)}\displaystyle\sum_{n=0}^\infty \frac{\mathcal{I}(\alpha,\zeta,n)}{\sqrt{\rho(n)}}\vert n \rangle~,
\end{equation} 
with the normalisation constant
\begin{equation}
\mathcal{N}^2(\alpha,\zeta)=\displaystyle\sum_{n=0}^\infty \frac{\vert \mathcal{I}(\alpha,\zeta,n)\vert ^2}{\rho(n)}~,
\end{equation}
where 
\begin{equation}
\rho(n)=\displaystyle\prod_{i=1}^n k(i) = \displaystyle\prod_{i=1}^n if^2(i), \qquad \rho(0)=1~.
\end{equation}
Inserting \textcolor{beamer@PRD}{(}\ref{Expansion}\textcolor{beamer@PRD}{)} into the eigenvalue equation \textcolor{beamer@PRD}{(}\ref{eigen}\textcolor{beamer@PRD}{)} replaced with the generalised ladder operators \textcolor{beamer@PRD}{(}\ref{GenLad}\textcolor{beamer@PRD}{)} or \textcolor{beamer@PRD}{(}\ref{qOsc}\textcolor{beamer@PRD}{)}, we end up with a three term recurrence relation
\begin{equation}\label{recurrence}
\mathcal{I}(\alpha,\zeta,n+1)=\alpha~\mathcal{I}(\alpha,\zeta,n)-\zeta ~k(n)~ \mathcal{I}(\alpha,\zeta,n-1),
\end{equation}
with $\mathcal{I}(\alpha,\zeta,0)=1$ and $\mathcal{I}(\alpha,\zeta,1)=\alpha$, which when solved, leads to the explicit form of the squeezed states for the models corresponding to the particular values of $k(n)$ \cite{Angelova_Hertz_Hussin}. Note that, the recurrence relation \textcolor{beamer@PRD}{(}\ref{recurrence}\textcolor{beamer@PRD}{)} may not be easy to solve, when one deals with the complicated choices of $k(n)$. However, it becomes fairly simple to work out, when one considers the special case $\zeta=0$, which corresponds to the coherent states. In this scenario, the recurrence relation is always solvable and the explicit form is well known 
\begin{equation}\label{NonLinear}
\vert \alpha\rangle=\frac{1}{\mathcal{N}(\alpha)}\displaystyle\sum_{n=0}^\infty \frac{\alpha^n}{\sqrt{\rho(n)}}\vert n \rangle~,
\end{equation}
which are also familiar as the nonlinear coherent states \cite{Filho_Vogel,Manko_Marmo_Solimeno_Zaccaria,Sivakumar}.
\end{section} 
\begin{section}{Squeezed states for the noncommutative harmonic oscillator}\label{sec3}
We will construct the squeezed states for the one dimensional harmonic oscillator \cite{Kempf,Bagchi_Fring,Dey_Fring_Gouba,Dey_Fring_squeezed}
\begin{equation}\label{NHHam}
H=\frac{P^2}{2m}+\frac{m\omega^2}{2}X^2-\hbar\omega\left(\frac{1}{2}+\frac{\tau}{4}\right)
\end{equation}
defined in the noncommutative space satisfying 
\begin{equation}\label{ncrep}
[X,P]=i\hbar\left(1+\check{\tau}P^2\right),\qquad X=\left(1+\check{\tau}p^2\right)x,\qquad P=p~.
\end{equation}
Here $\check{\tau}=\tau/(m\omega\hbar)$ has the dimension of inverse squared momentum with $\tau$ being dimensionless. The noncommutative observables $X$, $P$ are represented in \textcolor{beamer@PRD}{(}\ref{ncrep}\textcolor{beamer@PRD}{)} in terms of the standard canonical variables $x$, $p$ satisfying $[x,p]=i\hbar$. The Hamiltonian $H$ is clearly non-Hermitian with respect to the standard inner product. However, one can easily construct the isospectral Hermitian counterpart of the non-Hermitian Hamiltonian \textcolor{beamer@PRD}{(}\ref{NHHam}\textcolor{beamer@PRD}{)}, when the Hamiltonian $H$ is assumed to be pseudo/quasi Hermitian \cite{Scholtz_Geyer_Hahne,Bender_Boettcher,Mostafazadeh_2002,Bender_Making_Sense}, i.e. the non-Hermitian Hamiltonian $H$ and the Hermitian Hamiltonian $h$ are related by a similarity transformation $h=\eta H\eta^{-1}$, with $\eta^\dagger\eta$ being a positive definite operator playing the role of the metric. The corresponding eigenstates $\vert\Phi\rangle$ and $\vert\phi\rangle$ of $H$ and $h$, respectively, are then simply related as $\vert\Phi\rangle=\eta^{-1}\phi$. The Dyson map $\eta$, whose adjoint action relates the non-Hermitian Hamiltonian \textcolor{beamer@PRD}{(}\ref{NHHam}\textcolor{beamer@PRD}{)} to its isospectral Hermitian counterpart $h$, is easily found to be $\eta=(1+\check{\tau}p^2)^{-1/2}$. With the help of the metric $\eta$ we compute \cite{Dey_Fring_squeezed}
\begin{equation}\label{HermHam}
h=\eta H \eta^{-1}=\frac{p^2}{2m}+\frac{m\omega^2}{2}x^2+\frac{\omega\tau}{4\hbar}\left(p^2x^2+x^2p^2+2xp^2x\right)-\hbar\omega\left(\frac{1}{2}+\frac{\tau}{4}\right)+\mathcal{O}(\tau^2)~.
\end{equation} 
We consider a perturbative treatment here and decompose the above Hamiltonian \textcolor{beamer@PRD}{(}\ref{HermHam}\textcolor{beamer@PRD}{)} as $h=h_0+h_1$. Now, taking $h_0$ to be the standard harmonic oscillator and following the common techniques of Rayleigh-Schr{\"o}dinger perturbation theory, the energy eigenvalues of $H$ and $h$ were computed \cite{Kempf, Dey_Fring_Gouba, Dey_Fring_squeezed} to lowest order to
\begin{equation}
E_n=\hbar\omega e_n=\hbar\omega (A n+B n^2)+\mathcal{O}(\tau^2)~,
\end{equation}
with $A=(1+\tau/2)$ and $B=\tau/2$. The eigenstates were calculated correspondingly to
\begin{equation}\label{eigenstates}
\vert\phi_n\rangle=\vert n\rangle-\frac{\tau}{16}\sqrt{(n-3)^{(4)}}\vert n-4\rangle +\frac{\tau}{16}\sqrt{(n+1)^{(4)}}\vert n+4\rangle +\mathcal{O}(\tau^2)~,
\end{equation} 
where $Q^{(n)}:=\prod_{k=0}^{n-1}(Q+k)$ denotes the Pochhammer symbol with the raising factorial. In what follows, we will drop the explicit mentioning of the order in $\tau$, understanding that all our computations are carried out to first order. Having obtained all the prerequisites, we can now construct the coherent states for the noncommutative harmonic oscillator \textcolor{beamer@PRD}{(}\ref{NHHam}\textcolor{beamer@PRD}{)}. The explicit form of which follows from \textcolor{beamer@PRD}{(}\ref{NonLinear}\textcolor{beamer@PRD}{)} as
\begin{equation}\label{3.6}
\vert \alpha\rangle = \frac{1}{\mathcal{N}(\alpha)}\displaystyle\sum_{n=0}^{\infty}\frac{\alpha^n}{\sqrt{\rho(n)}}\vert \phi_n\rangle,
\end{equation}
which, when replaced by \textcolor{beamer@PRD}{(}\ref{eigenstates}\textcolor{beamer@PRD}{)}, we obtain
\begin{alignat}{1}
\vert \alpha\rangle &= \frac{1}{\mathcal{N}(\alpha)}\Bigg[\displaystyle\sum_{n=0}^{\infty} \frac{\alpha^n}{\sqrt{\rho(n)}}\left(\vert n\rangle+\frac{\tau}{16}\sqrt{\frac{(n+4)!}{n!}}\vert n+4\rangle\right)-\frac{\tau}{16}\displaystyle\sum_{n=4}^{\infty} \frac{\alpha^n}{\sqrt{\rho(n)}}\sqrt{\frac{n!}{(n-4)!}}\vert n-4\rangle\Bigg] \label{3.7}\\
&= \frac{1}{\mathcal{N}(\alpha)}\Bigg[\displaystyle\sum_{n=0}^{\infty} \frac{\alpha^n}{\sqrt{\rho(n)}}\left(1-\frac{\tau}{16}\alpha^4\frac{f(n)!}{f(n+4)!}\right)\vert n \rangle+\frac{\tau}{16}\displaystyle\sum_{n=0}^{\infty} \frac{\alpha^n}{\sqrt{\rho(n)}}\sqrt{\frac{(n+4)!}{n!}}\vert n+4\rangle\Bigg] \label{3.8}\\
&= \frac{1}{\mathcal{N}(\alpha)}\Bigg[\displaystyle\sum_{n=0}^{\infty} \frac{\alpha^n}{\sqrt{\rho(n)}}\left(1-\frac{\tau}{16}\alpha^4\frac{f(n)!}{f(n+4)!}\right)\vert n \rangle+\frac{\tau}{16}\displaystyle\sum_{n=4}^{\infty} \frac{\alpha^{n-4}}{\sqrt{\rho(n)}}\frac{n!}{(n-4)!}\frac{f(n)!}{f(n-4)!}\vert n\rangle\Bigg] \label{3.9}\\
&= \frac{1}{\mathcal{N}(\alpha)}\displaystyle\sum_{n=0}^{\infty} \frac{\mathcal{C}(\alpha,n)}{\sqrt{\rho(n)}}\vert n\rangle~,\label{3.10}
\end{alignat}
where 
\begin{equation}\notag
\mathcal{C}(\alpha,n)=\left\{ \begin{array}{lcl}
\alpha^n-\frac{\tau}{16}\alpha^{n+4}\frac{f(n)!}{f(n+4)!} & \mbox{if}
& 0\leq n \leq 3 \\ \alpha^n-\frac{\tau}{16}\alpha^{n+4}\frac{f(n)!}{f(n+4)!}+\frac{\tau}{16}\alpha^{n-4}\frac{n!}{(n-4)!}\frac{f(n)!}{f(n-4)!}\ & \mbox{if} & n\geq 4~.\end{array}\right.
\end{equation} 
Whereas, the squeezed states for the noncommutative harmonic oscillator \textcolor{beamer@PRD}{(}\ref{NHHam}\textcolor{beamer@PRD}{)} are obtained by solving the recurrence relation \textcolor{beamer@PRD}{(}\ref{recurrence}\textcolor{beamer@PRD}{)} with $k(n)=An+Bn^2$. The solution is obtained in terms of the Gauss hypergeometric function $_2F_1$ as follows
\begin{equation}\label{SqueezedState}
\mathcal{I}(\alpha,\zeta,n)=i^n\left(\zeta B\right)^{n/2}\left(1+\frac{A}{B}\right)^{(n)}~_2F_1\Bigg[-n,\frac{1}{2}+\frac{A}{2B}+\frac{i\alpha}{2\sqrt{\zeta B}};1+\frac{A}{B};2\Bigg]~,
\end{equation}
where $Q^{(n)}$ denotes the Pochhammer symbol as mentioned after \textcolor{beamer@PRD}{(}\ref{eigenstates}\textcolor{beamer@PRD}{)}. Following similar steps as \textcolor{beamer@PRD}{(}\ref{3.6}\textcolor{beamer@PRD}{)}-\textcolor{beamer@PRD}{(}\ref{3.10}\textcolor{beamer@PRD}{)}, and, by replacing equation \textcolor{beamer@PRD}{(}\ref{SqueezedState}\textcolor{beamer@PRD}{)} into \textcolor{beamer@PRD}{(}\ref{Expansion}\textcolor{beamer@PRD}{)} along with $\rho(n)=k(n)!=[nf^2(n)]!=e_n!=[n(A+Bn)]!$, we obtain the explicit form of the squeezed states of the noncommutative harmonic oscillator as follows
\begin{equation}\label{SqStateNC}
\vert \alpha,\zeta\rangle =\frac{1}{\mathcal{N}(\alpha,\zeta)}\displaystyle\sum_{n=0}^{\infty} \frac{\mathcal{S}(\alpha,\zeta,n)}{\sqrt{\rho(n)}}\vert n\rangle~,
\end{equation}
where 
\begin{equation}\notag
\mathcal{S}(\alpha,\zeta,n)=\left\{ \begin{array}{lcl}
\mathcal{I}(\alpha,\zeta,n)-\frac{\tau}{16}\frac{f(n)!}{f(n+4)!}\mathcal{I}(\alpha,\zeta,n+4) & \mbox{if}
& 0\leq n \leq 3 \\ \mathcal{I}(\alpha,\zeta,n)-\frac{\tau}{16}\frac{f(n)!}{f(n+4)!}\mathcal{I}(\alpha,\zeta,n+4)+\frac{\tau}{16}\frac{n!}{(n-4)!}\frac{f(n)!}{f(n-4)!}\mathcal{I}(\alpha,\zeta,n-4) & \mbox{if} & n\geq 4~.\end{array}\right.
\end{equation} 
In the harmonic oscillator limit $\tau=0$, i.e. $f(n)=1$, we obtain the reduced form, which is the squeezed states of the ordinary harmonic oscillator precisely
\begin{equation}\label{SqStateHO}
\vert \alpha,\zeta\rangle_{ho}^{}=\frac{1}{\mathcal{N}(\alpha,\zeta)}\displaystyle\sum_{n=0}^\infty \frac{1}{\sqrt{n!}}\left(\frac{\zeta}{2}\right)^{n/2}\mathcal{H}_n(\frac{\alpha}{\sqrt{2\zeta}})\vert n \rangle~,
\end{equation}
where $\mathcal{H}_n(\alpha)$ denote the Hermite polynomials.
\end{section}
\begin{section}{Measure of entanglement with quantum beam splitter}\label{sec4}
A beam splitter is a well known optical interferometer, which has two input and two output ports. The lights passing through the input ports are partly reflected and partly transmitted with the amplitude reflection and transmission coefficients being $r$ and $t$, respectively. The quantum version of the classical beam splitter is obtained by replacing the incoming electromagnetic fields with a set of annihilation operators $a$ and $b$ corresponding to two different inputs \cite{Gerry_Knight_Book}. The output fields are, then, realized with the unitarily transformed operators $c=\mathcal{B}a \mathcal{B}^\dagger$ and $d=\mathcal{B}b \mathcal{B}^\dagger$, such that 
\begin{equation}\label{OutputComm}
[c,c^\dagger]=1 \qquad \text{and} \qquad [d,d^\dagger]=1.
\end{equation}
The unitary operator $\mathcal{B}$ is known as the beam splitter operator
\begin{equation}
\mathcal{B}=e^{\frac{\theta}{2}(a^\dagger b e^{i\phi}-a b^\dagger e^{-i\phi})}~,
\end{equation}
where $\theta$ denotes the angle of the beam splitter and $\phi$ is the phase difference between the reflected and transmitted fields. The conditions \textcolor{beamer@PRD}{(}\ref{OutputComm}\textcolor{beamer@PRD}{)} impose the restriction on the reflection and transmission amplitudes, $\vert r\vert^2+\vert t\vert^2=1$, with $r=-e^{-i\phi}\sin(\theta/2)$ and $t=\cos(\theta/2)$. For a $50:50$ beam splitter, $r$ and $t$ are naturally equal in amplitude, $\vert r\vert=\vert t\vert=1/\sqrt{2}$. The effect of the beam splitter operator on a bipartite input state composed of a usual Fock state $\vert n\rangle$ at one of the inputs and a vacuum state $\vert 0\rangle$ at the other, is fairly well known  \cite{Kim_Son_Buzek_Knight}
\begin{equation}\label{BeamFock}
\mathcal{B}\vert n\rangle_a \vert 0\rangle_b=\displaystyle\sum_{q=0}^n \begin{pmatrix}
n \\ q
\end{pmatrix}^{1/2}t^q r^{n-q}~\vert q\rangle_c\vert n-q\rangle_d~.
\end{equation}
However, the outcome becomes much more complicated when one considers the squeezed states \cite{Hertz_Hussin_Eleuch}, especially in the noncommutative space $\vert \alpha,\zeta\rangle$ \textcolor{beamer@PRD}{(}\ref{Expansion}\textcolor{beamer@PRD}{)}. The output state in this case is computed to
\begin{equation}
\vert \text{out}\rangle =\mathcal{B}\vert \alpha,\zeta\rangle_a\vert 0\rangle_b=\frac{1}{\mathcal{N}(\alpha,\zeta)}\displaystyle\sum_{n=0}^\infty \frac{\mathcal{S}(\alpha,\zeta,n)}{\sqrt{k(n)!}}\mathcal{B}\vert n\rangle_a\vert 0\rangle_b~,
\end{equation}
which when replaced by equation \textcolor{beamer@PRD}{(}\ref{BeamFock}\textcolor{beamer@PRD}{)}, achieves the form
\begin{eqnarray}
\vert \text{out}\rangle &=& \frac{1}{\mathcal{N}(\alpha,\zeta)}\displaystyle\sum_{n=0}^\infty\displaystyle\sum_{q=0}^n \frac{\mathcal{S}(\alpha,\zeta,n)}{\sqrt{k(n)!}}\begin{pmatrix}
n \notag \\ q
\end{pmatrix}^{1/2}t^q r^{n-q}~\vert q\rangle_c\vert n-q\rangle_d \\
&=& \frac{1}{\mathcal{N}(\alpha,\zeta)}\displaystyle\sum_{q=0}^\infty\displaystyle\sum_{m=0}^{\infty-q} \frac{\mathcal{S}(\alpha,\zeta,m+q)}{\sqrt{m!q!}f(m+q)!}t^q r^m~\vert q\rangle_c\vert m\rangle_d~. \label{output}
\end{eqnarray}
The normalisation constant is evaluated from the requirement, $\langle \text{out}\vert \text{out}\rangle =1$ as follows
\begin{equation}
\mathcal{N}^2(\alpha,\zeta)=\displaystyle\sum_{q=0}^\infty\displaystyle\sum_{m=0}^{\infty-q} \frac{\vert \mathcal{S}(\alpha,\zeta,m+q)\vert^2}{m!q! f^2(m+q)!}\vert t\vert^{2q} \vert r\vert^{2m}.
\end{equation} 
The most exciting feature of a quantum beam splitter is that it produces entangled output states, if at least one of the input fields is nonclassical \cite{Kim_Son_Buzek_Knight,Xiang-Bin}. It is well known that one does not obtain the entangled states in the output ports, when one transmits coherent states through the input ports \cite{Kim_Son_Buzek_Knight}. The reason behind this is that the coherent states are classical in nature \cite{Dey_Fring_PRA,Dey_Fring_squeezed,Dey_Fring_Gouba_Castro}, whereas the squeezed states are highly nonclassical \cite{Walls,Loudon_Knight} and very useful for the creation of the entangled states. For further analysis on the nonclassical behaviours of the states, albeit in noncommutative space, one may look at the recent work by one of the authors \cite{Dey}. 
\begin{subsection}{Linear entropy}
In this section we will analyse the entanglement produced by the squeezed states in the noncommutative space, that we have constructed in section \ref{sec2} and \ref{sec3}. There are many options available for the measurement of quantum entanglement. For instance, one could quantify logarithmic negativity \cite{Vidal_Werner}, entanglement distillation \cite{Kwiat1}, concurrence \cite{Rungta_Buzek_Caves_Hilary_Milburn}, squashed / CMI entanglement \cite{Christandl}, relative entropy \cite{Vedral}, von-Neumann entropy \cite{Schumacher} etc. However, they are not so easy to compute for complicated systems like noncommutative oscillators. Although, von-Neumann entropy seems to be straightforward to calculate, once the density matrix $\rho$ is known
\begin{equation}
S(\rho)=-\it{Tr}(\rho\log_2\rho)= -\sum_{x}\lambda_x\log_2\lambda_x~.
\end{equation}
Nevertheless, one requires to diagonalise the matrix to compute the eigenvalues $\lambda_x$, which is not always possible, specially for difficult systems like us. A relatively easy way would be to compute the linear entropy  
\begin{equation}
S=1-\it{Tr}(\rho_\mathcal{A}^2),
\end{equation}
where $\rho_\mathcal{A}^{}$ is the reduced density operator of the system $\mathcal{A}$, which is obtained by performing partial trace over the system $\mathcal{D}$ of the density operator $\rho_{\mathcal{A}\mathcal{D}}^{}$. The value of $S$ varies in between $0$ and $1$, where $0$ corresponds to the case of the density matrix of pure states and $1$ correlates to the case of completely mixed states producing the maximum entanglement. Let us first calculate the density matrix for the case at hand \textcolor{beamer@PRD}{(}\ref{output}\textcolor{beamer@PRD}{)}
\begin{eqnarray}
\rho_{\mathcal{A}\mathcal{D}}^{} &=& \vert \text{out}\rangle\langle\text{out}\vert \notag \\
&=& \frac{1}{\mathcal{N}^2(\alpha,\zeta)}\displaystyle\sum_{q=0}^\infty\displaystyle\sum_{s=0}^\infty\displaystyle\sum_{m=0}^{\infty-q}\displaystyle\sum_{n=0}^{\infty-s} \frac{\mathcal{S}(\alpha,\zeta,m+q)\mathcal{S}^\ast(\alpha,\zeta,n+s)}{\sqrt{m!q!n!s!}f(m+q)!f(n+s)!}t^q\overline{t^s} r^m\overline{r^n}~\vert q,m\rangle\langle s,n\vert~,
\end{eqnarray}
where $\vert q,m\rangle\equiv\vert q\rangle_c \vert m\rangle_d$, such that the partial trace is computed to
\begin{equation}
\rho_\mathcal{A}^{}=\frac{1}{\mathcal{N}^2(\alpha,\zeta)}\displaystyle\sum_{q=0}^\infty\displaystyle\sum_{s=0}^\infty\displaystyle\sum_{m=0}^{\infty-\text{max}(q,s)} \frac{\mathcal{S}(\alpha,\zeta,m+q)\mathcal{S}^\ast(\alpha,\zeta,m+s)}{m!\sqrt{q!s!}f(m+q)!f(m+s)!}t^q\overline{t^s} \vert r\vert^{2m}~\vert q\rangle\langle s\vert~.
\end{equation} 
Therefore the linear entropy becomes
\begin{eqnarray}
S &=& 1-\frac{1}{\mathcal{N}^4(\alpha,\zeta)}\displaystyle\sum_{q=0}^\infty\displaystyle\sum_{s=0}^\infty\displaystyle\sum_{m=0}^{\infty-\text{max}(q,s)} \displaystyle\sum_{n=0}^{\infty-\text{max}(q,s)} \vert t\vert^{2(q+s)}\vert r\vert^{2(m+n)} \notag \\ 
&& \times\frac{\mathcal{S}(\alpha,\zeta,m+q)\mathcal{S}^\ast(\alpha,\zeta,m+s)\mathcal{S}(\alpha,\zeta,n+s)\mathcal{S}^\ast(\alpha,\zeta,n+q)}{q!s!m!n!f(m+q)!f(m+s)!f(n+s)!f(n+q)!}~.\label{Entropy}
\end{eqnarray}
This is evident that the analytical study of the entropy \textcolor{beamer@PRD}{(}\ref{Entropy}\textcolor{beamer@PRD}{)} is quite difficult here. Even the numerical treatment is very challenging indeed, when large number of energy levels are considered. One needs to be very careful when one deals with the entropy with finite number of energy levels, because the energy spectrum of the original model is infinite. We are very concerned with the fact and every time we have investigated the minimum requirement of the number of energy levels for which the series converges for the particular values of the parameters that we have chosen in our computations. A $50:50$ beam splitter has been utilised here, such that $\theta$ is taken to be equal to $\pi/2$. However, the entropy \textcolor{beamer@PRD}{(}\ref{Entropy}\textcolor{beamer@PRD}{)} depends on the square of the absolute value of the phase difference between the reflected and transmitted fields, so that $\phi$ becomes entirely irrelevant to our case.

Our investigations are carried out mainly in two different directions. First, we compute the linear entropy of the output states while one of the input states of the beam splitter is a coherent state in noncommutative space \textcolor{beamer@PRD}{(}\ref{3.10}\textcolor{beamer@PRD}{)}, as depicted in figure \ref{fig1}. Second, we study mutual comparisons of the entanglement of the noncommutative squeezed states \textcolor{beamer@PRD}{(}\ref{SqStateNC}\textcolor{beamer@PRD}{)} input with that of the squeezed state of the ordinary harmonic oscillators \textcolor{beamer@PRD}{(}\ref{SqStateHO}\textcolor{beamer@PRD}{)}. The outcomes for different values of the squeezing parameters have been demonstrated in figure \ref{fig2}. The linear entropies of the noncommutative oscillators are much higher than the usual ones for all values of $\alpha$, which indicate that the output states coming out of the squeezed states of the noncommutative harmonic oscillator are more entangled than that of the squeezed states of ordinary harmonic oscillators.
\begin{figure}[h]
\centering   \includegraphics[width=7.6cm,height=5.7cm]{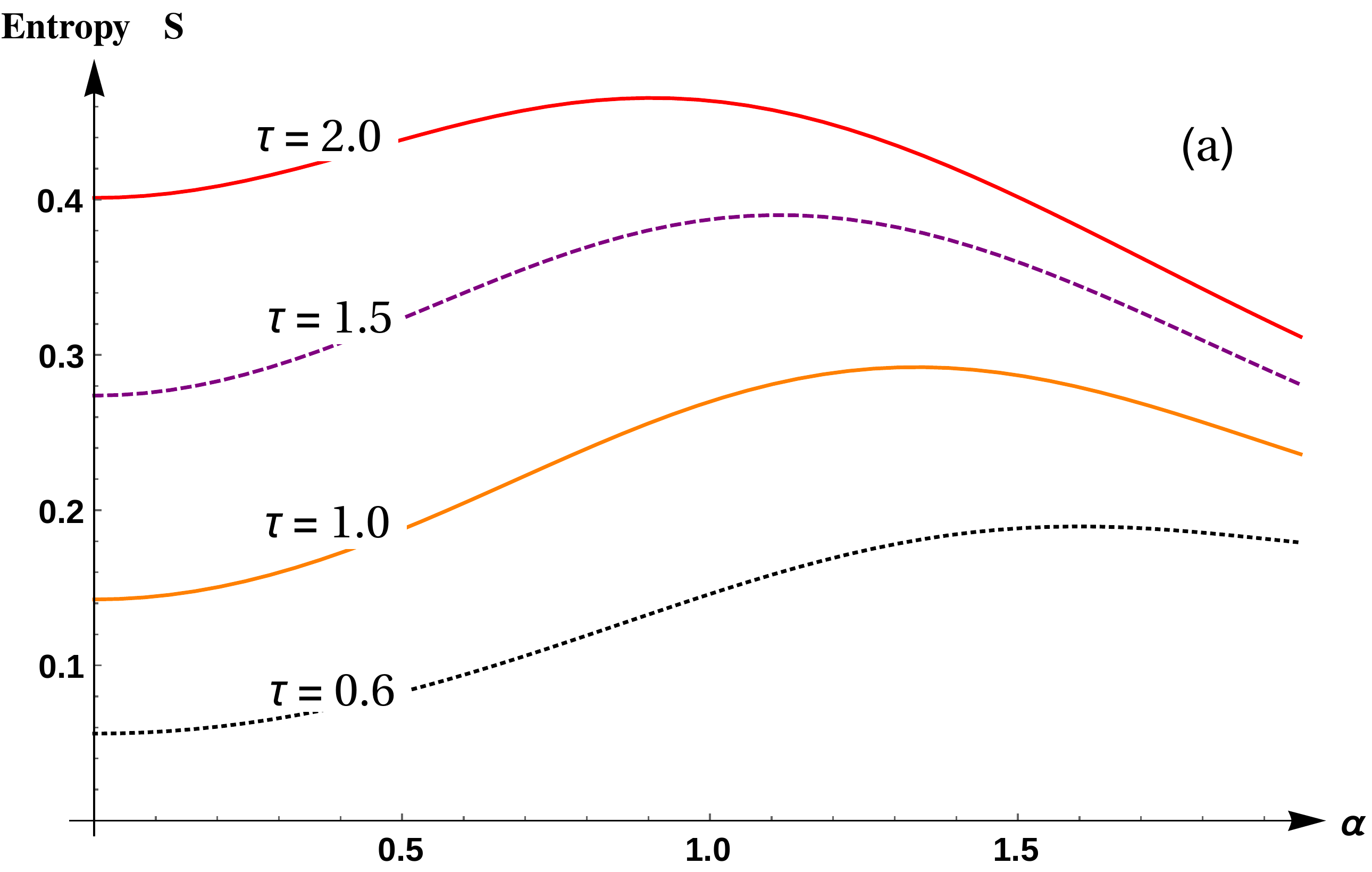}
\includegraphics[width=7.6cm,height=5.7cm]{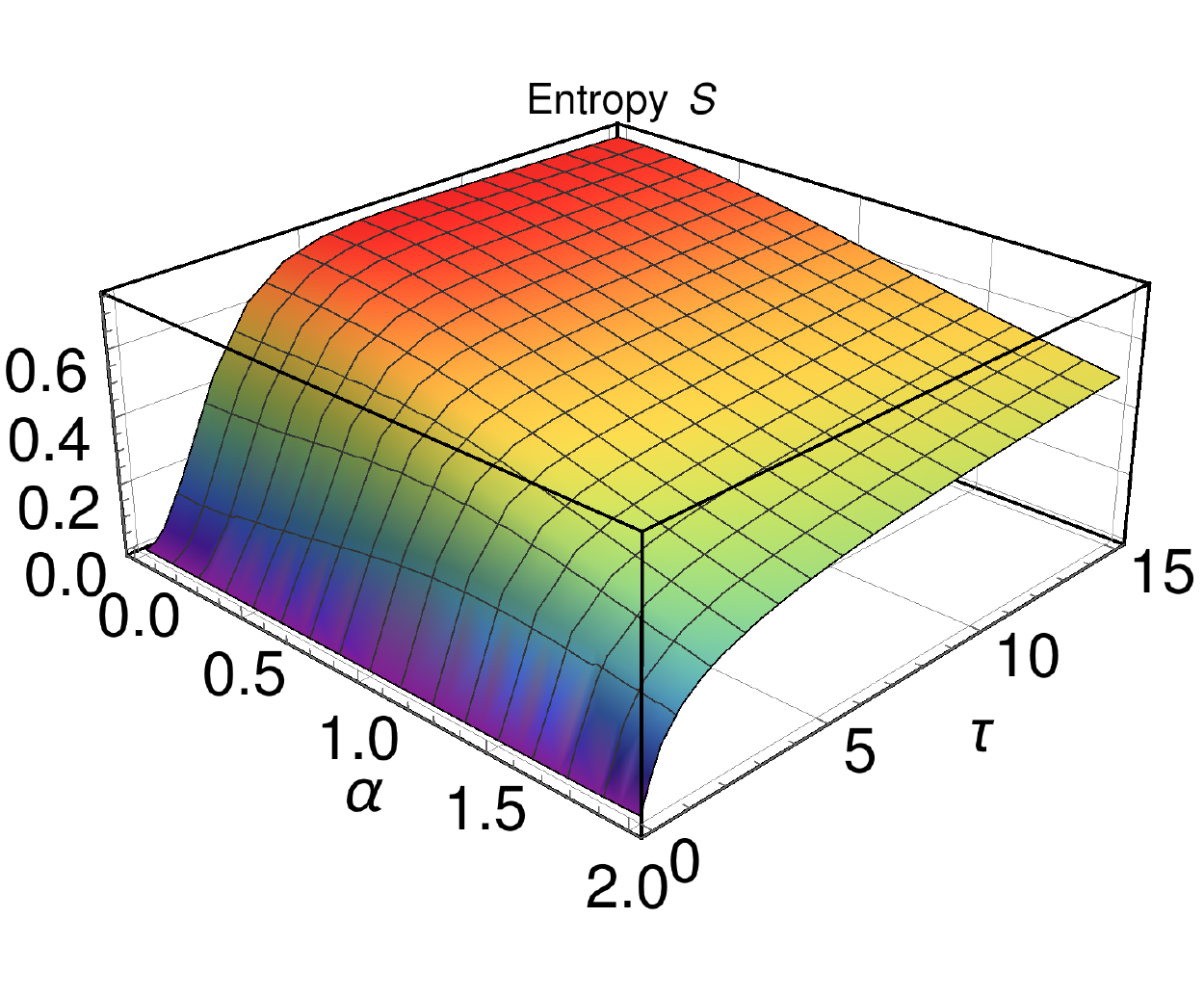}
\includegraphics[width=1.0cm,height=5.7cm]{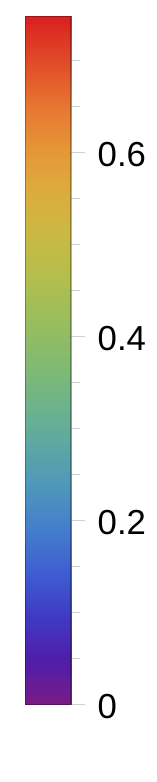}
\caption{\small{Linear entropy for the noncommutative coherent state input (a) as function of $\alpha$ for different values of $\tau$ (b) as functions of $\alpha$ and $\tau$. Number of energy levels considered $=20$ in each case.}}
\label{fig1}
\end{figure}
\begin{figure}[H]
\centering   \includegraphics[width=8.2cm,height=6.0cm]{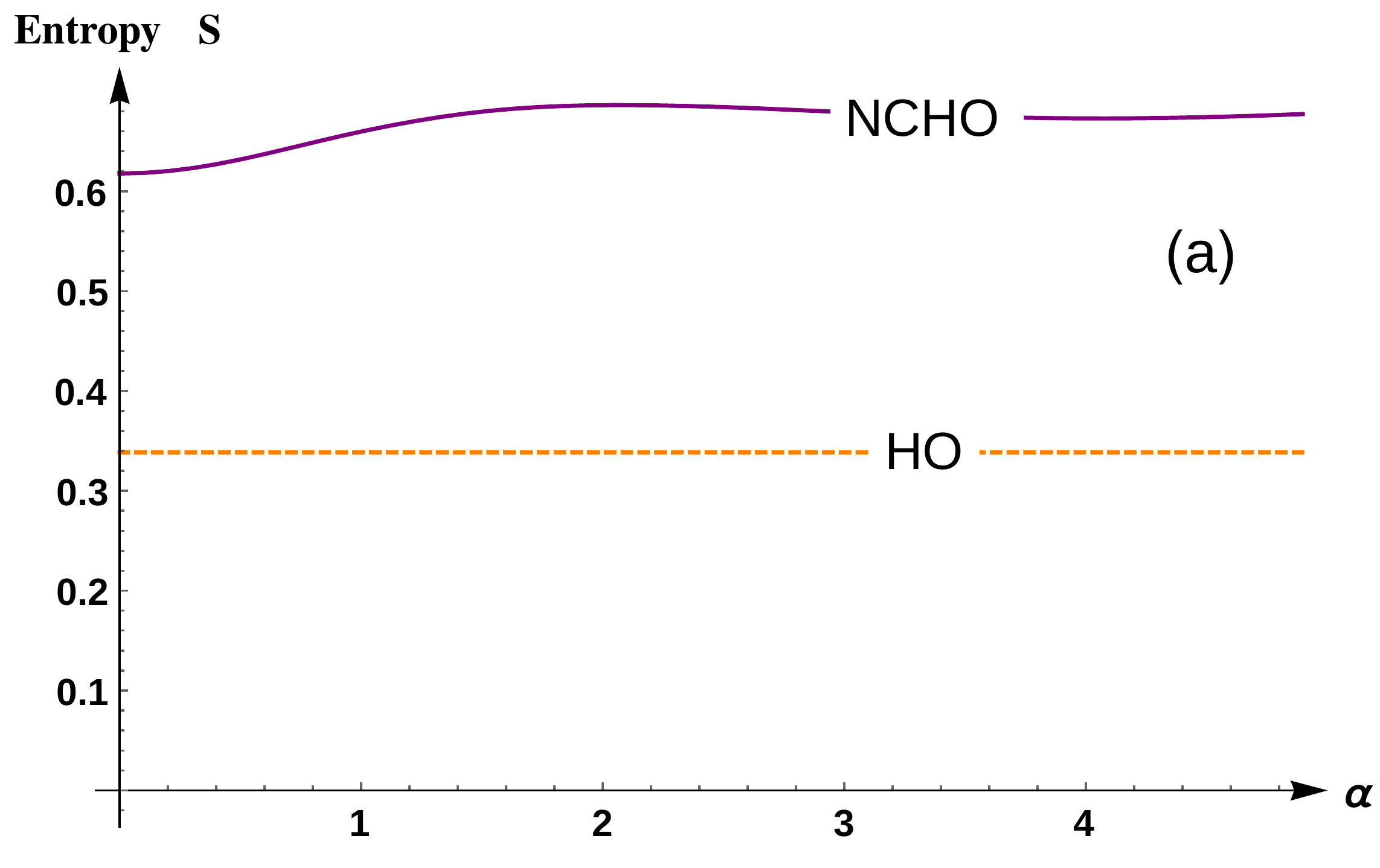}
\includegraphics[width=8.2cm,height=6.0cm]{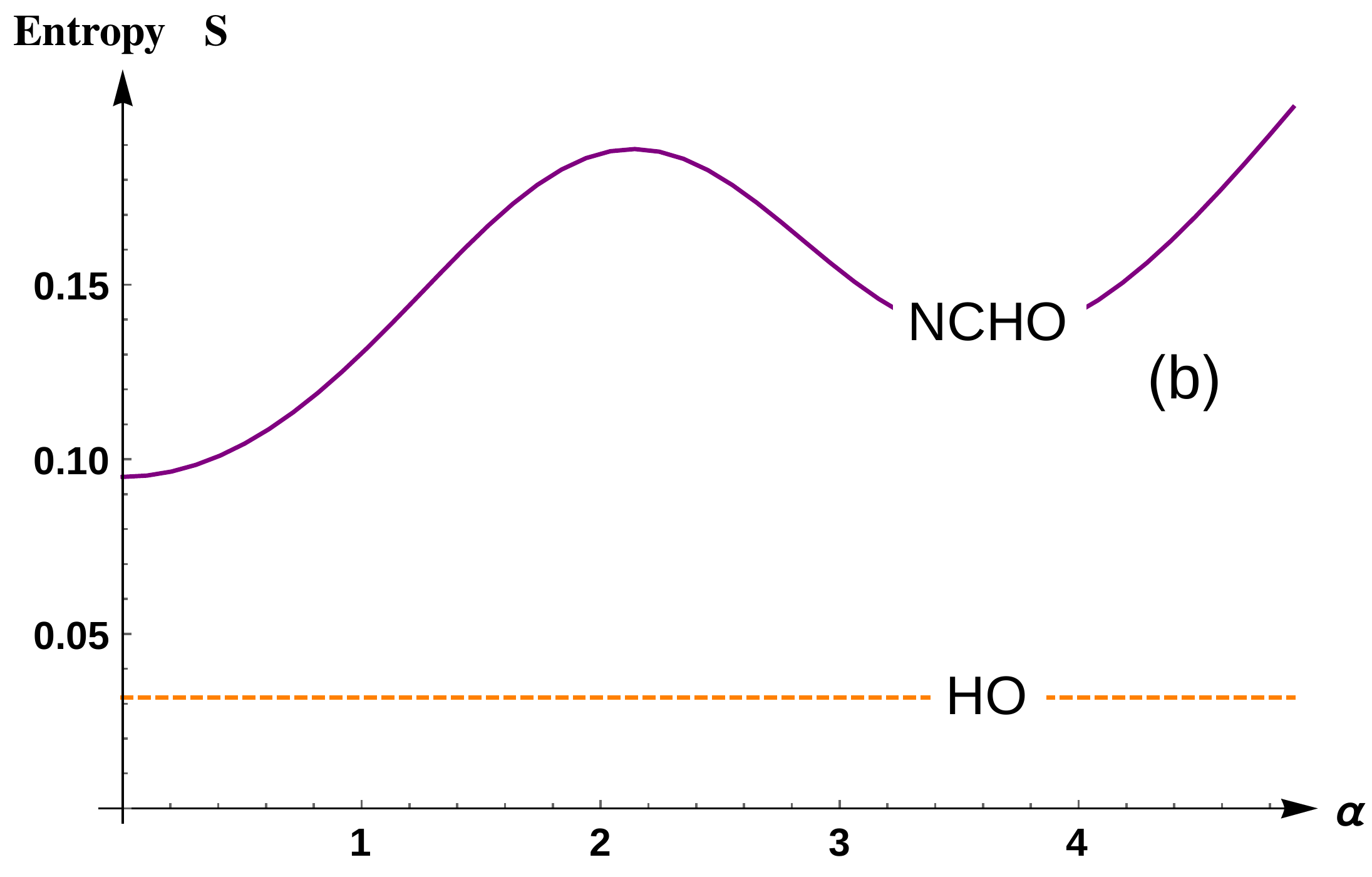}
\caption{\small{Linear entropy for the noncommutative squeezed state (solid, purple) versus squeezed state of ordinary harmonic oscillator (dashed, orange) as function of $\alpha$ for $\tau=0.5$ (a) $\zeta=0.75$ (b) $\zeta=0.25$. Number of energy levels considered = 40 in each case.}}
\label{fig2}
\end{figure}
\noindent On the other hand, when we consider the coherent states in noncommutative space in one of the inputs, we see that little amount of entropies are obtained in the output as shown in figure \ref{fig1}(a). Therefore, the behaviour of the noncommutative coherent states are more interesting than the coherent states of the ordinary harmonic oscillator, where we do not obtain the entangled output states at all. With reference to figure \ref{fig1}, we suggest that the noncommutative coherent states are slightly nonclassical in nature. While, the classical like manner of the noncommutative states have been investigated by many authors \cite{Ghosh_Roy,Dey_Fring_squeezed,Dey_Fring_Gouba_Castro,Dey}. The dual nature of the coherent states in noncommutative spaces have also been found earlier, rather based on the analytical treatment; see, for instance, \cite{Dey_Fring_squeezed,Roy_Roy}. However, for quite obvious reasons, the output states are always more entangled, when the input states are squeezed states, rather than the coherent states in noncommutative space. 
\begin{figure}[H]
\centering   \includegraphics[width=10.0cm,height=7.0cm]{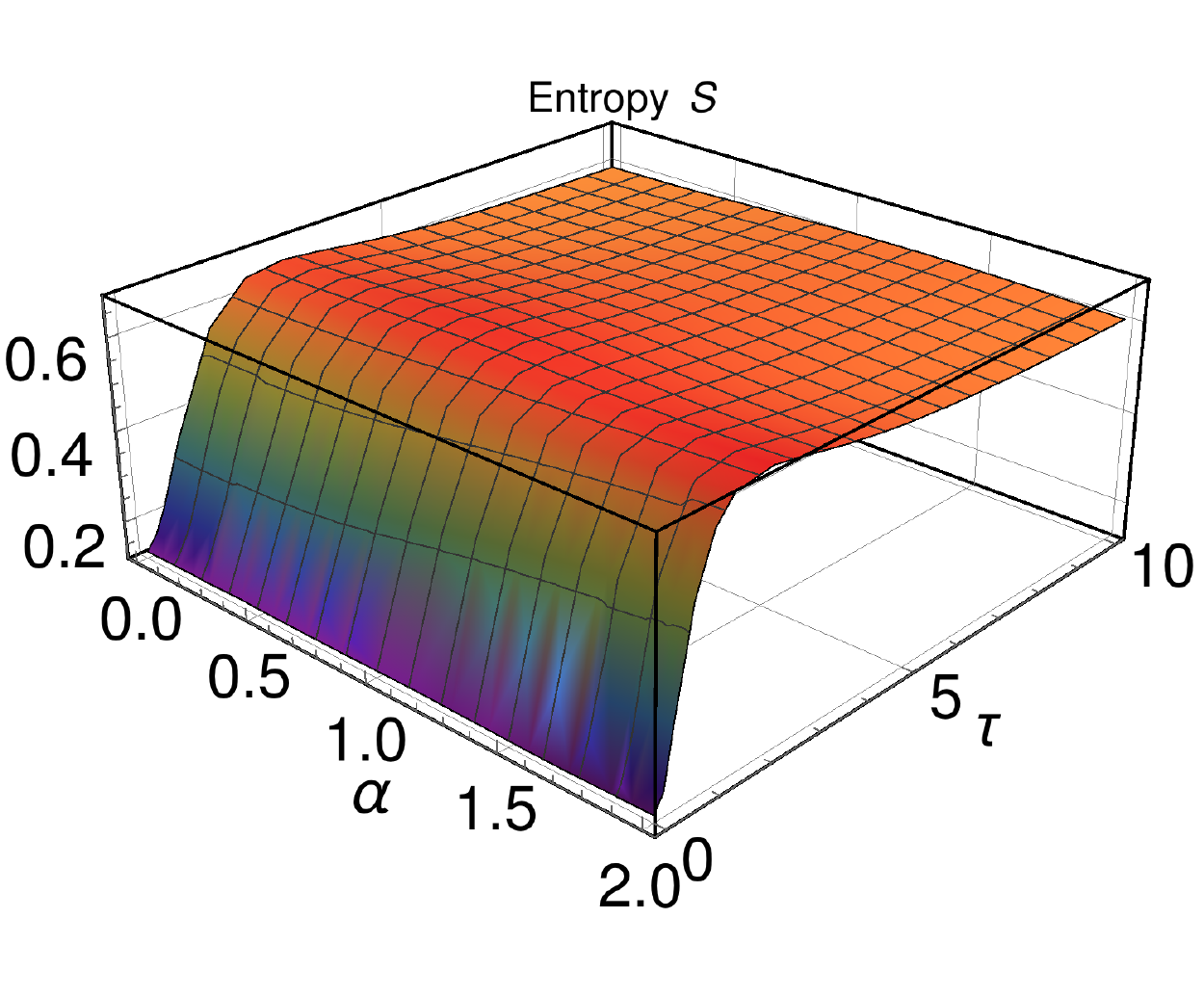}
\includegraphics[width=1.2cm,height=7.0cm]{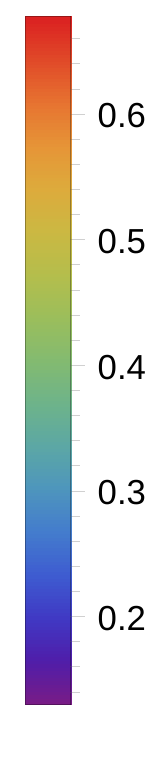}
\caption{\small{Linear entropy for the noncommutative squeezed state input as functions of $\alpha$ and $\tau$ for $\zeta=0.5$. Number of energy levels considered = 10.}}
\label{fig3}
\end{figure} 
\noindent The most exciting effect is that the key role on the behaviour of the linear entropy is played by the noncommutative parameter $\tau$, which is quite obvious in figure \ref{fig3}. The value of the entropy for the noncommutative case coincides with the entropy of the ordinary harmonic oscillator, when $\tau=0$, which is expected. However, the entropy increases rapidly with the increase of the value of $\tau$ and saturates at sufficiently high value, irrespective of all values of $\alpha$. The behaviour is detected in all of the two scenarios that we are discussing here. First, for the case of the coherent states, which has been depicted in figure \ref{fig1}(b). Second, for the squeezed states, which is plotted in figure \ref{fig3}. Certainly, as per our expectation, we obtain better result in the latter case. The above analysis indicates that one may utilize the noncommutative systems to improve the entanglement further the usual quantum mechanical systems.
\end{subsection} 
\end{section}
\begin{section}{Conclusions}\label{sec5}
In the present work, we have constructed the entangled states on a noncommutative space leading to a generalized version of Heisenberg's uncertainty relation. A lossless $50:50$ quantum mechanical beam splitter has been utilised for this purpose. We provide an explicit expression of the nonclassical state; i.e. squeezed state of the perturbative noncommutative harmonic oscillator, which is itself very difficult to construct. However, our principal result is to demonstrate the superiorities of utilising the noncommutative spaces over the usual quantum mechanical systems for the purpose of creating entangled states. We have shown that the noncommutative space provides an extra degree of freedom by which one can increase the degree of entanglement beyond the ordinary systems. Depending on the value of the noncommutative parameter $\tau$, one produces higher entangled states in noncommutative spaces than the usual systems. An additional observation is to point out the dual nature of the coherent states in noncommutative space. Noncommutative coherent states exhibit both classical-like nature as well as nonclassicality, which we believe to be a very interesting finding in this context.

There are various directions in which our analysis might be taken forward. For instance, one may study the entanglement for many other models in the same space or in some other noncommutative spaces. More challenging is the construction of such entangled states in higher dimensions. Apart from using quantum beam splitter, constructing quantum entanglement of the noncommutative models using some other techniques would be insightful indeed. 

\vspace{0.5cm} \noindent \textbf{\large{Acknowledgements:}} S. D. is supported by the Post Doctoral Fellowship jointly funded by the Laboratory of Mathematical Physics of the Centre de Recherches Math{\'e}matiques (CRM) and by Prof. Syed Twareque Ali, Prof. Marco Bertola and Prof. V{\'e}ronique Hussin. The authors thank Prof. Joris Van der Jeugt for his useful discussions.

\end{section}


\end{document}